\documentclass{kluwer_mod}    

\usepackage[dvips]{graphicx}

\newdisplay{guess}{Conjecture}

\begin{document}                                                                                   
\begin{article}
\begin{opening}         
\title{Galaxy Evolution tool: Construction and Applications} 
\author{Yeshe \surname{Fenner} and Brad K. \surname{Gibson}} 
\institute{Centre for Astrophysics \& Supercomputing, Swinburne University,
Australia}  \author{Marco \surname{Limongi}} 
\institute{Osservatorio Astronomico di Roma}

\runningauthor{Yeshe Fenner}
\runningtitle{Galaxy Evolution tool: Construction and Applications}



\end{opening}            

Galaxy Evolution tool (GEtool) is a software package currently being developed
to self-consistently model the chemical and spectral evolution of disk
galaxies.  GEtool will soon be available to the community through a
web-based interface\footnote{http://astronomy.swin.edu.au/GEtool/}
that will enable users to predict observable properties of model galaxies such
as colours, spectral gradients, Lick indices and elemental abundances. We
present an application of the code to a dual accretion phase model of the Milky
Way in order to assess the role of massive stars in the chemical composition of
the Galaxy. A updated set of stellar yields, covering a range of stellar initial
masses and metallicities ($13 < m / M_{\odot} < 80$ and $Z/Z_{\odot} = 0,
10^{-3}, 1$), has recently been calculated by Limongi et al (2001, in prep)
(LSC01). This poster presents results from the first chemical
evolution model to incorporate these new stellar yields.

\begin{figure} 
\tabcapfont
\centerline{
\begin{tabular}{p{8cm} p{8cm}} 
\includegraphics[width=8cm]{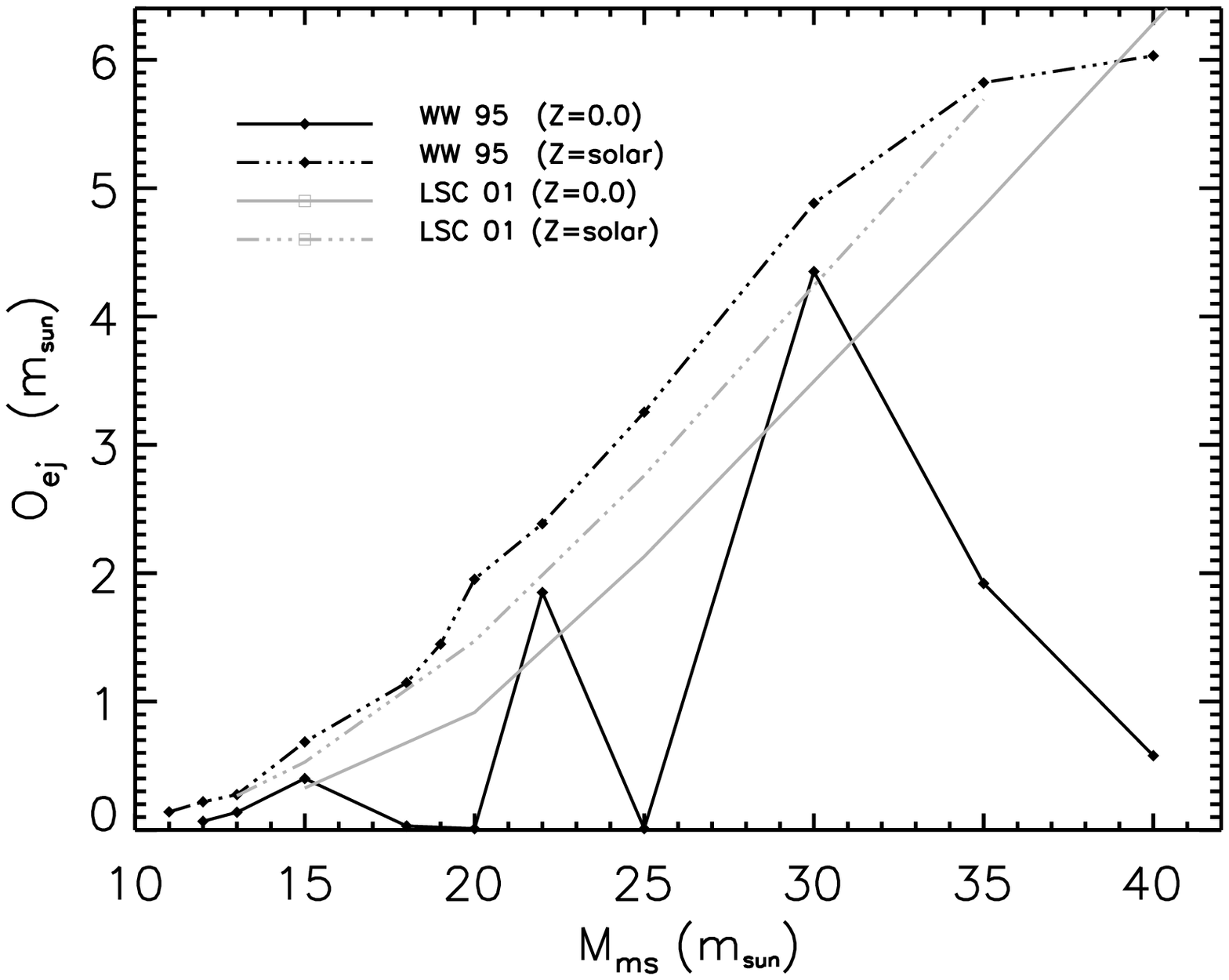} &
\includegraphics[width=8cm]{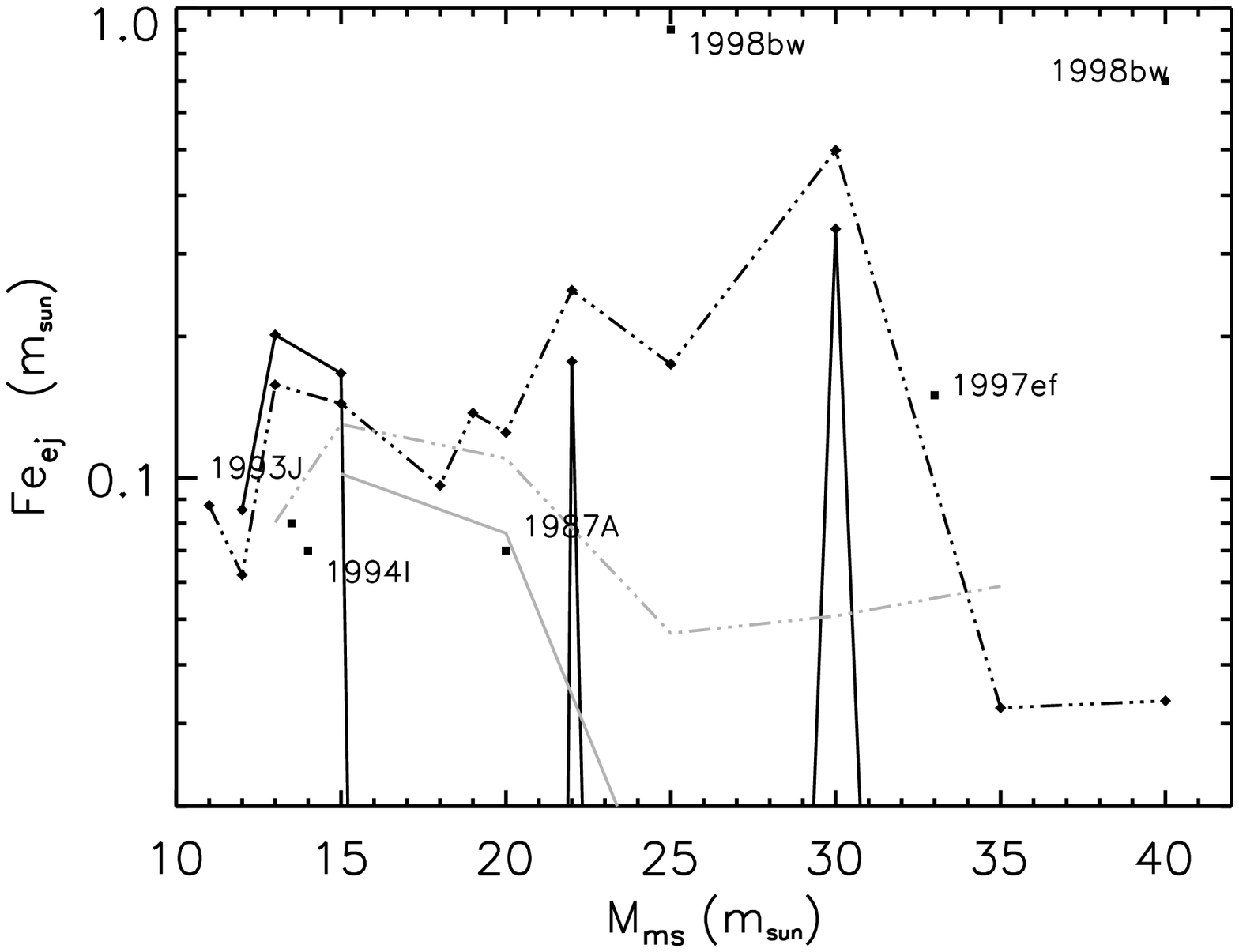} \\
\end{tabular}}
\caption{The ejected mass of O (\emph{left}) and Fe (\emph{right}) as a
function of main sequence mass from LSC01 (\emph{grey lines}) and WW95
(\emph{black lines}). The estimated mass of iron ejected from five observed SNe
are shown. Two models have been proposed for SN1998bw.}
\label{Fevsm} 
\end{figure}

\begin{figure} 
\tabcapfont
\centerline{
\begin{tabular}{p{8cm} p{8cm}} 
\includegraphics[width=8cm]{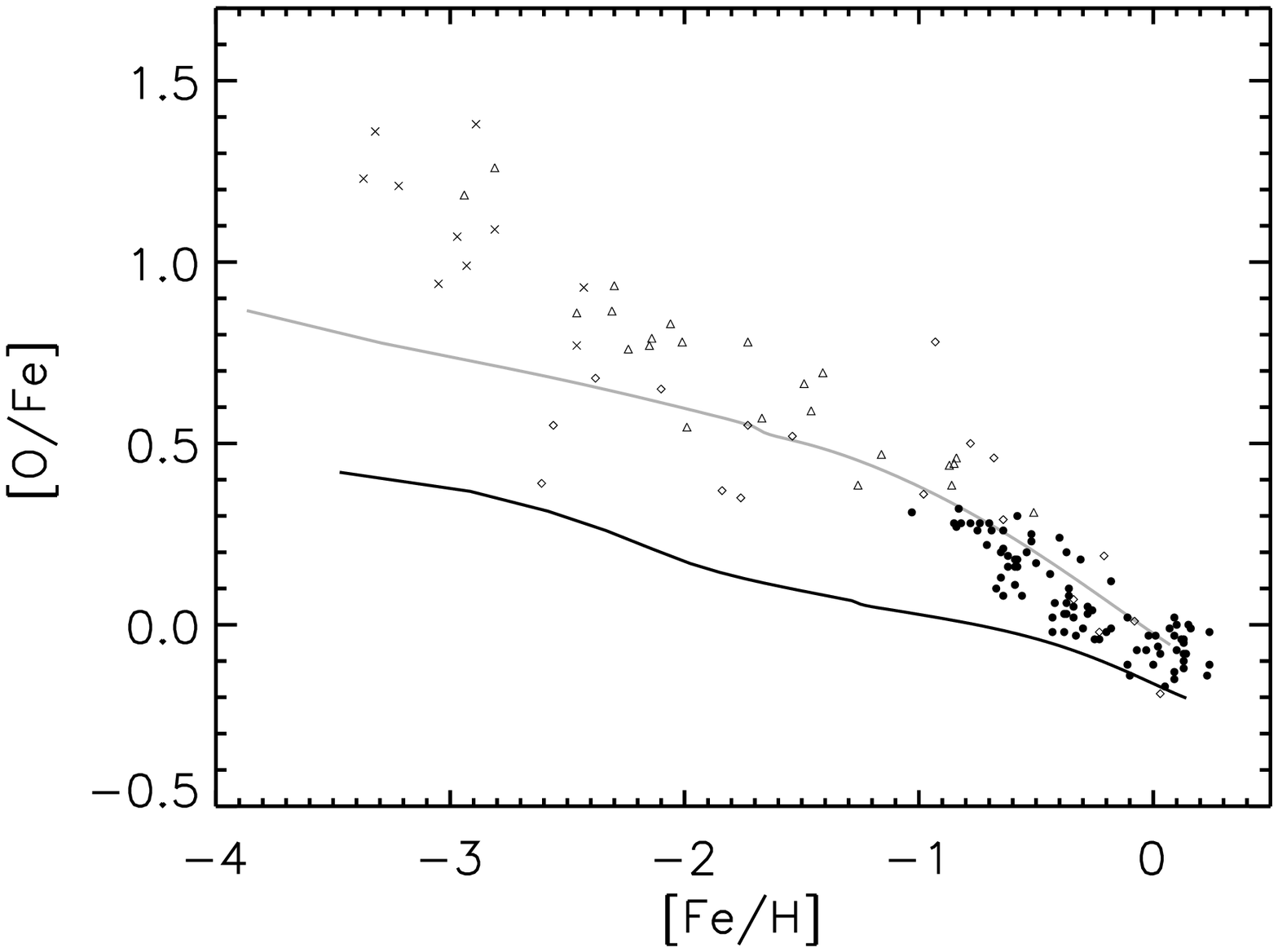} &
\includegraphics[width=8cm]{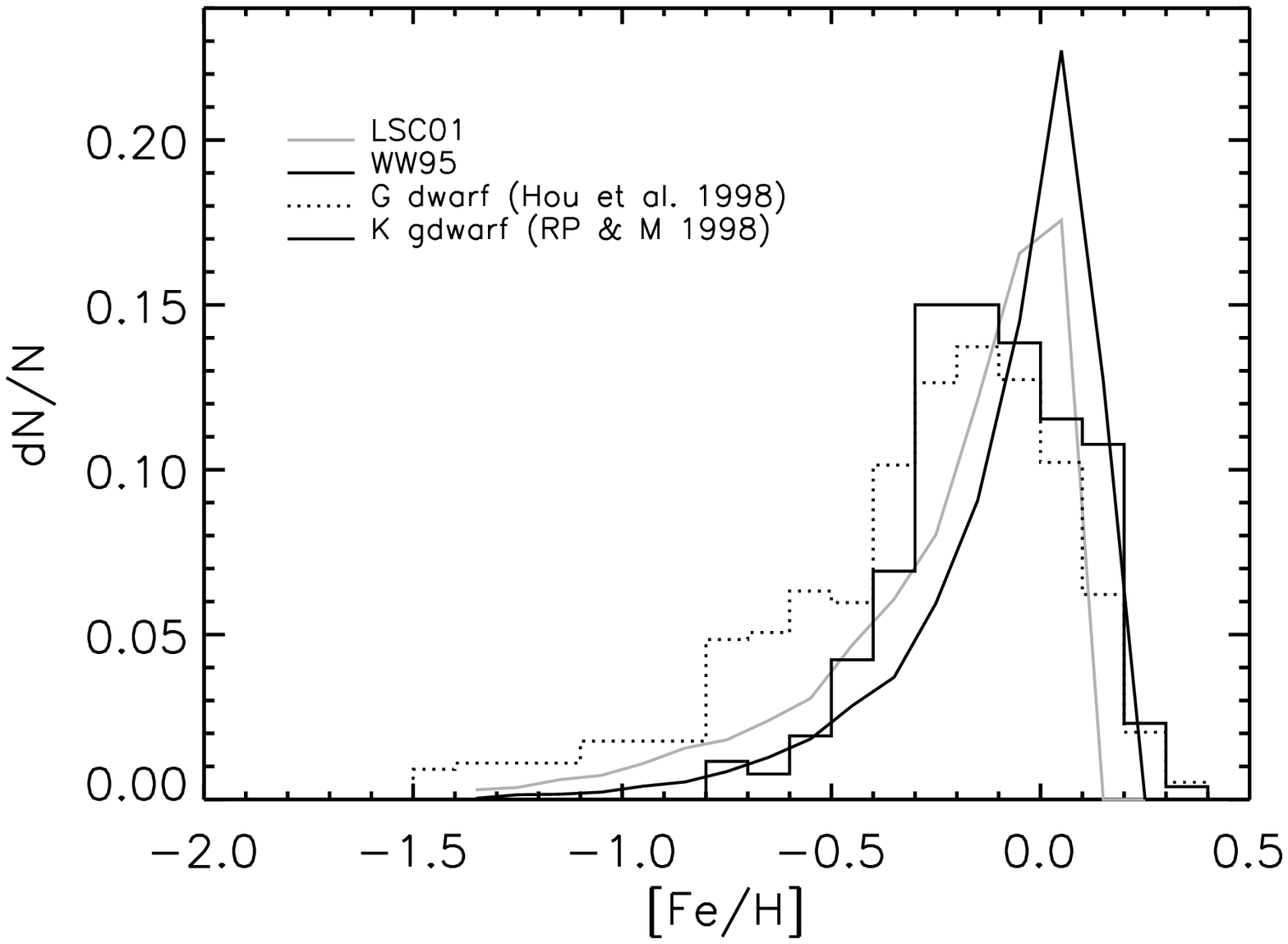} \\
\end{tabular}}
\centerline{
\begin{tabular}{@{\hspace{4cm}} p{8cm}@{\hspace{0cm}} p{8cm}} 
a. &  b. \\
\end{tabular}}
\caption{\emph{a}. O/Fe vs metallicity for stars are plotted for
comparison with predictions from dual-infall chemical evolution models. Grey
lines represent models using LSC01 yields while black lines depict results
from WW95. Observational data are from Edvardsson et al. (1993, A\&AS, 102,
603: \emph{filled circles}), Israelian et al. (2001, ApJ, 551, 833:
\emph{crosses}), Boesgaard  et al.(1999, ApJ, 117, 492: \emph{triangles}) and
Carretta et al. (2000, A\&A, 356, 238: \emph{diamonds}) \hspace{1cm} \emph{b}.
The predicted G dwarf distribution from the two stellar yield models (\emph{grey and
black lines}) are shown against the observed metallicity distribution of G and K
dwarfs in the solar neighbourhood (\emph{histograms}).}  \label{OonFe}  
\end{figure}

The age-metallicity relation, G-dwarf distribution and evolution
of abundance ratios predicted using the sets of yields from
 LSC01 and Woosley \& Weaver (1995, ApJS, 101, 181) (WW95) are compared with
observations. Stellar yields are one of the most important ingredients in
Galactic chemical evolution models, yet the ejected mass of iron-peak elements
is very uncertain due to the edge of the iron core being so close to the mass
cut. Further uncertainties are introduced by the possibility that much of the
synthesised iron in higher mass stars falls back to core. Since there are few
supernova observations with which to directly infer the iron ejecta, we
compared the predicted and observed trend of key abundance patterns to
indirectly constrain iron yields. We found that the trend of [O/Fe] and [Mg/Fe]
vs [Fe/H] inferred from recent measurements cannot be recovered if the iron
yield of massive stars increases with initial mass, as in WW95 model. Instead,
the observational constraints are better satisfied if we assume the iron yield
decreases for initial masses greater than $\sim$ 25 $M_{\odot}$ (see
Figure~\ref{Fevsm} and Figure~\ref{OonFe}.a.). Both models
shown in Figure~\ref{OonFe}.b. reproduce the observed
G-dwarf distribution in the solar neighbourhood.

\end{article}
\end{document}